\documentstyle[psfig,aps,preprint]{revtex}

\def\beq{\begin{equation}}
\def\eeq{\end{equation}}
\def\bea{\begin{eqnarray}}
\def\eea{\end{eqnarray}}
\newcommand{\sla}{\!\!\!\!/ \,}

\begin{document}

\draft

\title{Equilibrium and Non-Equilibrium Hard Thermal Loop Resummation 
in the Real Time Formalism\footnote{Supported by BMBF, GSI Darmstadt, and DFG}}
\author{Margaret E. Carrington$^1$, Hou Defu$^2$, and Markus H. 
Thoma$^{3,4}$\footnote{Heisenberg fellow}}
\address{$^1$Department of Physics, Brandon University, Brandon, Manitoba, Canada R7A 6A9 \\
$^2$Institut f\"ur Theoretische Physik, Universit\"at
Regensburg, D-93040 Regensburg, Germany\\
$^3$Institut f\"ur Theoretische Physik, Universit\"at Giessen,
D-35392 Giessen, Germany\\
$^4$European Center for Theoretical Studies in Nuclear Physics and Related
Areas, Villa Tambosi, Strada delle Tabarelle 286, I-38050 Villazzano, Italy}
\date{\today}
\maketitle

\begin{abstract}
We investigate the use of the hard thermal loop (HTL) resummation technique in 
non-equilibrium field theory.  We use the Keldysh representation of the real 
time formalism (RTF).  We derive the HTL photon self 
energy and the resummed photon propagator. We show that no pinch singularities
appear in the non-equilibrium HTL effective propagator.  We discuss a possible
regularization mechanism 
for these singularities at higher orders. 
As an example of the application 
of the HTL resummation method within the RTF we discuss the damping rate 
of a hard electron. \end{abstract}


\section{Introduction}

Perturbation theory for gauge theories at finite temperature suffers from
infrared singularities and gauge dependent results for physical quantities.
These problems are avoided by using an effective perturbation theory
(Braaten-Pisarski method \cite{ref1}) which is based on the resummation
of hard thermal loop (HTL) diagrams into effective Green functions. This
powerful method was derived within the imaginary time formalism (ITF).
Using resummed Green functions, medium effects of the heat bath, such as
Debye screening, collective plasma modes, and Landau damping, are taken
into account. The HTL resummation technique has been applied to a number of 
interesting problems, in particular to the prediction of signatures and
properties of a quark-gluon plasma (QGP) expected to be produced in 
relativistic heavy ion collisions (for a review see \cite{ref2}). 

However, the use of thermal field theories for describing a QGP in
nucleus-nucleus collisions is restricted by the fact that at least the
early stage of such a collision leads to a fireball, which is not in
equilibrium. It is not clear if a complete thermal and chemical equilibrium will be
achieved later on. Hence, non-equilibrium effects in a 
parton gas should be considered for predicting signatures of
QGP formation and for obtaining a consistent picture of the fireball.
This can be done in the case of a chemically non-equilibrated parton gas by
means of rate equations \cite{ref3} or more generally by using
transport models \cite{ref4}. However, these approaches are based
on a semiclassical approximation. In particular, infrared divergences have
to be removed phenomenologically. Therefore it is desirable to derive
a Green function approach including medium effects as in the case of
the HTL resummation. For this purpose one has to abandon the ITF, which
is restricted to equilibrium situations. The real time formalism (RTF), on 
the other hand, can be extended to investigate non-equilibrium systems
\cite{ref5,ref6}.

The RTF involves choosing a contour in the complex energy plane which fulfills
the Kubo-Martin-Schwinger boundary condition and contains the real axis
\cite{ref5}. This leads to propagators and self energies which are given
by $2\times 2$ matrices. The choice of the contour is not unique.
We will
adopt the Keldysh or closed time path contour, which was invented
for the non-equilibrium case \cite{ref5}. In particular, we will demonstrate
the usefulness of the Keldysh representation \cite{ref7} based on
advanced and retarded propagators and self energies and show how potentially
dangerous terms (pinch singularities) \cite{ref8} in non-equilibrium are 
treated easily within this representation.

In the next section we review the Keldysh representation.  In section 3, we 
discuss the equilibrium calculation.  We consider QED and give the results of 
the real time calculation, in the HTL approximation, of the photon self energy, the
resummed photon propagator, and the electron damping rate. The results are, of 
course, identical to those of the ITF, which demonstrates that although the HTL 
resummation scheme was derived within the ITF, the result is independent of the 
choice of contour.  
In section 4, we extend the HTL resummation technique to 
off-equilibrium situations by following the equilibrium calculations outlined in section 3.  
We show that no pinch singularities
appear in the non-equilibrium HTL effective propagator. 

 \section{Keldysh representation}

In this section we review the Keldysh representation of the RTF.  The bare 
propagator for bosons reads \cite{ref6}
\beq
D(K)=\left (\begin{array}{cc} \frac{1}{K^2-m^2+i\epsilon} & 0\\
                              0 & \frac{-1}{K^2-m^2-i\epsilon}\\
            \end{array} \right ) 
-2\pi i\, \delta (K^2-m^2)\> \left (\begin{array}{cc}
n_B(k_0) & \theta (-k_0)+n_B(k_0)\\
\theta (k_0)+n_B(k_0) & n_B(k_0) \\ \end{array} \right ),
\label{e1} 
\eeq
where $K=(k_0,{\bf k})$, $k=|{\bf k}|$, $\theta $ denotes the step
function, and the distribution function is given by $n_B(k_0)=
1/[\exp(|k_0|/T)-1]$ in the equilibrium case.  

For fermions the bare propagator can be written as 
\bea
S(K)=(K \sla +m)\> && \left [ \left (\begin{array}{cc} 
\frac{1}{K^2-m^2+i\epsilon} & 0\\
0 & \frac{-1}{K^2-m^2-i\epsilon}\\
                          \end{array} \right )\right .\nonumber \\ 
&& +2\pi i\, \delta (K^2-m^2)\> \left .\left (\begin{array}{cc}
n_F(k_0) & -\theta (-k_0)+n_F(k_0)\\
-\theta (k_0)+n_F(k_0) & n_F(k_0) \\ \end{array} \right )\right ],
\label{e2} 
\eea
where the Fermi distribution is given by $n_F(k_0)=1/[\exp(|k_0|)+1]$
in equilibrium. 
The components of these propagators are not independent, but fulfill
the relation
\beq
G_{11}-G_{12}-G_{21}+G_{22}=0,
\label{e3}
\eeq
where $G$ stands for $D$ or $S$.

By an orthogonal transformation of these $2\times 2$ matrices we arrive
at a representation of the propagators in terms of advanced and retarded
propagators which was first introduced by Keldysh \cite{ref7}. The three independent
components of this representation are defined as \cite{ref6}
\bea
G_R & = & G_{11}-G_{12},\nonumber \\
G_A & = & G_{11}-G_{21},\nonumber \\
G_F & = & G_{11}+G_{22}.
\label{e4}
\eea
The inverted relations read
\bea
G_{11} & = & \frac{1}{2}\, (G_F+G_A+G_R),\nonumber \\
G_{12} & = & \frac{1}{2}\, (G_F+G_A-G_R),\nonumber \\
G_{21} & = & \frac{1}{2}\, (G_F-G_A+G_R),\nonumber \\
G_{22} & = & \frac{1}{2}\, (G_F-G_A-G_R).
\label{e5}
\eea

Similar relations to (\ref{e3}) and (\ref{e4}) hold for the self
energies \cite{ref10}:
\beq
\Pi_{11}+\Pi_{12}+\Pi_{21}+\Pi_{22}=0
\label{e9}
\eeq
and
\bea
\Pi_R & = & \Pi_{11}+\Pi_{12},\nonumber \\
\Pi_A & = & \Pi_{11}+\Pi_{21},\nonumber \\
\Pi_F & = & \Pi_{11}+\Pi_{22},
\label{e10}
\eea
where $\Pi $ stands for the self energy of a boson or fermion.

Using (\ref{e1}) and (\ref{e2}) in (\ref{e4}) the bare propagators of
the Keldysh representation are given by
\bea
D_R(K) & = & \frac{1}{K^2-m^2+i\, \mbox{sgn}(k_0) \epsilon},\nonumber \\
D_A(K) & = & \frac{1}{K^2-m^2-i\, \mbox{sgn}(k_0) \epsilon},\nonumber \\
D_F(K) & = & -2\pi i\, [1+2n_B(k_0)]\, \delta (K^2-m^2) 
\label{e6}
\eea
for bosons and
\bea
S_R(K) & = & \frac{K\sla +m}{K^2-m^2+i\, \mbox{sgn}(k_0) \epsilon},\nonumber \\
S_A(K) & = & \frac{K\sla +m}{K^2-m^2-i\, \mbox{sgn}(k_0) \epsilon},\nonumber \\
S_F(K) & = & -2\pi i\, (K\sla +m)\, [1-2n_F(k_0)]\, \delta (K^2-m^2)
\label{e7}
\eea
for fermions.
The bare propagators $D_F$ and $S_F$ can be written also as
\bea
D_F(K) & = & [1+2n_B(k_0)]\, \mbox{sgn}(k_0)\, [D_R(K)-D_A(K)],\nonumber \\
S_F(K) & = & [1-2n_F(k_0)]\, \mbox{sgn}(k_0)\, [S_R(K)-S_A(K)].
\label{e8}
\eea
In the non-equilibrium case, all of these equations are valid, with the equilibrium 
distribution functions ($n_B,\,\,n_F$) replaced by  non-equilibrium distribution 
functions ($f_B,\,\,f_F$) which depend on the four momentum and the 
space-time coordinate \cite{ref6}. 

Now we consider the situation for full (resummed) propagators.  In equilibrium, 
(\ref{e8}) is valid for full propagators as a consequence of the dissipation-fluctuation 
theorem \cite{ref10}. The polarization tensor satisfies, 
\beq
\Pi_F(K)=[1+2n_B(k_0)]\, \mbox{sgn}(k_0)\, [\Pi _R(K)-\Pi _A(K)]
\label{e10a}
\eeq
for bosons, and for fermions we have to replace $n_B$ by $-n_F$.  
Out of equilibrium however, the situation is more complicated.  
Equations (\ref{e8}) and (\ref{e10a}) are not satisfied by resummed 
propagators out of equilibrium.  Additional terms occur which appear 
to give rise to pinch singularities.  In section 4 we will discuss 
these terms in detail. 

\section{Equilibrium}

In this section we consider the hot QED plasma in equilibrium.  We discuss 
the HTL resummation technique in the context of the Keldysh representation 
of the RTF, as a starting point for our study of non-equilibrium situations.  

\subsection{HTL photon self energy}

The first step of the Braaten-Pisarski method is to extract the HTL
diagrams which have to be resummed into effective Green functions. A
typical example is the HTL photon self energy.
It is given by the diagram of fig.1, where the
momenta of the internal electron lines are of the order of the temperature 
or larger.
Applying standard Feynman rules one finds 
\beq
\Pi^{\mu\nu}(P)=-ie^2\int \frac {d^4K}{(2\pi )^4} tr \left [\gamma ^\mu 
S(Q)\gamma^ \nu S(K)\right ],
\label{e11}
\eeq
where $S$ denotes the electron propagator and $Q=K-P$. 
The retarded self energy is defined in
(\ref{e10}), 
\bea 
&& \Pi _R^{\mu \nu }(P)=\Pi _{11}^{\mu \nu }(P)+\Pi _{12}^{\mu \nu }(P) 
\nonumber \\
&& =-ie^2\int \frac {d^4K}{(2\pi )^4}\left \{ tr\left [\gamma ^\mu S_{11}(Q)
\gamma ^\nu S_{11}(K)\right ]-tr\left [\gamma ^\mu S_{21}(Q)\gamma ^\nu
S_{12}(K)\right ]\right \},
\label{e12}
\eea
where the minus sign in front of the second term comes from the vertex
of the type 2 fields \cite{ref5}. In the following we will neglect the 
electron mass assuming $m\ll T$ and write the electron propagator as
$S_{ij}(K)\equiv K\sla \tilde \Delta _{ij}(K)$. For now we 
will restrict 
ourselves to the longitudinal component of the self energy $\Pi ^L\equiv
\Pi ^{00}$. Performing the trace over the 
$\gamma $-matrices  
and using  (\ref{e5}) gives, 
\bea
\Pi _R^L(P)=-2ie^2\int \frac{d^4K}{(2\pi )^4} (q_0k_0+{\bf q}\cdot {\bf k})
&& \biggl [\tilde \Delta _F(Q)\tilde \Delta _R(K)+\tilde \Delta _A(Q)
\tilde \Delta _F(K)\nonumber \\
&& +\tilde \Delta _A(Q)\tilde \Delta _A(K)+\tilde 
\Delta _R(Q)\tilde \Delta _R(K)\biggr ].
\label{e14}
\eea
Terms proportional to $(\tilde \Delta _F(Q))^2$ that contain products of 
$\delta $-functions, which might cause
pinch singularities \cite{ref5}, do not appear.  This cancellation is well 
established in equilibrium calculations.    A great advantage of the Keldysh 
representation is that 
the cancellation is immediately evident, before any momentum integrals are done.  

To proceed further we do the integral using bare electron propagators and taking the HTL 
approximation. This approximation  is based on the assumption that we can
distinguish between soft momenta of the order $eT$ and hard ones of the order $T$,
which is possible in the weak coupling limit $e\ll 1$. We assume that the
external momentum $P$ is soft (because it is only for soft momenta that the HTL
self energies have to be resummed), and that the internal momentum
$K$ is hard.\footnote{In the ITF, i.e. in euclidean space, this assumption 
corresponds to $|p_0|, p\ll k$ \cite{ref2}. In the RTF (Minkowski space) however,
the requirement $|P|\ll k$ is sufficient since the exact one-loop self energies
coincide with the HTL ones on the light cone $P^2=0$ \cite{ref11b}.}  The resulting 
integral  can be done
analytically and gives the final result:
\beq
\Pi_R^L(P)=-3 m_\gamma^2 \left (1-\frac{p_0}{2 p}\ln \frac{p_0+p+i\epsilon}
{p_0-p+i\epsilon} \right),
\label{e17}
\eeq
where $m_\gamma=eT/3$ is the effective photon mass.
This result agrees with the result  in the ITF \cite{ref1,ref2}
(found earlier by Weldon and Klimov using the high temperature 
approximation \cite{ref11c}, which is equivalent to the HTL limit \cite{ref2}).
Analogously one obtains for the advanced photon self energy
\bea
\Pi_A^L(P) & = & \Pi _{11}^L(P)+\Pi _{21}^L(P)\nonumber \\
& = & -3 m_\gamma^2 \left (1-\frac{p_0}{2 p}\ln \frac{p_0+p-i\epsilon}
{p_0-p-i\epsilon} \right).
\label{e18}
\eea
The transverse part of the HTL photon self energy, $\Pi _T(P)=(\delta _{ij}-
p_ip_j/p^2)\Pi _{ij}(P)/2$, is computed in a similar way yielding
\beq 
\Pi_{R,A}^{T}(P)=\frac{3}{2}\, m_\gamma^2\, \frac{p_0^2}{p^2}
\left[ 1- \left( 1-\frac{p^2}{p_0^2} \right) \frac{p_0}{2 p}\ln
\frac{p_0+p\pm i\epsilon}{p_0-p\pm i\epsilon} \right].
\label{e20}
\eeq

Next we calculate $\Pi _F^L=-\Pi _{12}^L-\Pi _{21}^L$ (see (\ref{e9})
and (\ref{e10})) within the HTL approximation. As we will show in section 4,
this quantity is necessary to obtain the resummed propagator out
of equilibrium. Using (\ref{e5}) we obtain
\bea
\Pi _F^L(P)=-2ie^2\int \frac{d^4k}{(2 \pi )^4}\, (q_0k_0+{\bf q}\cdot
{\bf k})\> && \{\tilde \Delta _F(Q)\tilde \Delta _F(K)-[\tilde \Delta _R(Q)
-\tilde \Delta _A(Q)]\nonumber \\ 
&& [\tilde \Delta _R(K)-\tilde \Delta _A(K)]\}.
\label{e38a}
\eea
Extracting $\tilde{\Delta}_F$ from (\ref{e7}), using $\tilde \Delta _R
(Q)-\tilde \Delta _A(Q)=-2\pi i\, \mbox{sgn}(q_0)\, \delta (Q^2)$,  and taking the 
HTL approximation we obtain, 
\beq
\Pi _F^L(P)=-\frac{4ie^2}{\pi p}\theta (p^2-p_0^2)\, \int _0^\infty
dk\, k^2\, n_F(k)\, [1-n_F(k)] =-6\pi i\, m_\gamma ^2 \frac{T}{p} \theta (p^2-p_0^2).
\label{e40}
\eeq
The transverse part is given analogously by
\beq
\Pi _F^T(P)=-3\pi i\, m_\gamma ^2 \frac{T}{p}\left (1-\frac {p_0^2}{p^2}\right ) 
\theta (p^2-p_0^2).
\label{e40a}
\eeq
Note that the HTL expression for $\Pi _F$ is of higher order in the
coupling constant than $\Pi _{R,A}$  for soft momenta $p\sim eT$.
This observation also follows directly from
(\ref{e10a}) for soft $k_0$.
It is easy to show that these HTL results 
satisfy (\ref{e10a}) for soft $p_0$.

\subsection{Resummed photon propagator}

The second step of the Braaten-Pisarski method is the construction of the
effective Green functions to be used in the effective perturbation theory.
The resummed photon propagator, for instance, describing the propagation of 
a collective plasma mode,
is given by the Dyson-Schwinger equation of fig.2, where we adopt
the HTL result for the photon self energy. The equation
reads in Coulomb gauge ($D^{00}\equiv D^L$)
\beq
{D^*}^L=D^L+D^L\Pi ^L{D^*}^L,
\label{e21}
\eeq
where the propagators and self energy are $2\times 2$ matrices and
$*$ indicates a resummed propagator and not a complex conjugation. 
Throughout this paper we use the Coulomb gauge, which is convenient for
later applications \cite{ref2}. Since the final results for physical 
quantities are gauge independent using the HTL resummation method, we may
choose any gauge.  

Using
the identities (\ref{e3}) for the bare and resummed propagators,
(\ref{e9}) for the self energies, and the definitions (\ref{e4}) for the
advanced and retarded propagators $D_{A,R}$ and $D^*_{A,R}$ it is easy 
to show that
\beq
{D^*}_{R,A}^L=D_{R,A}^L+D_{R,A}^L\Pi _{R,A}^L{D^*}_{R,A}^L.
\label{e22}
\eeq 
$\,$From this expression we find for the effective longitudinal retarded and 
advanced photon propagators
\beq
{D^*}_{R,A}^L(P)=\left [p^2+3m_\gamma ^2\left (1-\frac {p_0}{2p}\ln
\frac{p_0+p\pm i\epsilon}{p_0-p\pm i\epsilon}\right )\right ]^{-1}.
\label{e23}
\eeq
$\,$From (\ref{e8}) we obtain, 
\beq
{D^*}_F^L(P)=[1+2n_B(p_0)]\, \mbox{sgn}(p_0)\, \left [{D^*}_R^L(P)-{D^*}_A^L(P)
\right ].
\label{e24}
\eeq
 Introducing
the spectral function \cite{ref11a}
\beq
\rho _L(P)\equiv -\frac{1}{\pi } Im {D^*}_R^L(P)
\label{e25}
\eeq
the propagator (\ref{e24}) can be written as
\beq
{D^*}_F^L(P)=-2\pi i\, [1+2n_B(p_0)]\, \mbox{sgn}(p_0)\, \rho _L(P).
\label{e26}
\eeq
Compared to the bare propagator we simply have to replace the bare
spectral function $\mbox{sgn}(p_0)\delta (P^2)$ in (\ref{e6}) by the spectral 
function for the effective propagator.

For the effective transverse photon propagator in Coulomb gauge we obtain
analogously 
\beq
{D^*}_{R,A}^T(P)=\left \{ p_0^2-p^2-\frac{3}{2}m_\gamma ^2\frac{p_0^2}{p^2} 
\left [1-\left (1-\frac {p^2}{p_0^2}\right ) \ln \frac{p_0+p\pm i\epsilon}
{p_0-p\pm i\epsilon}\right ]\right \} ^{-1}
\label{e27}
\eeq
and
\beq
{D^*}_F^T(P)=-2\pi i\, [1+2n_B(p_0)]\, \mbox{sgn}(p_0)\, \rho _T(P)
\label{e28}
\eeq
with the transverse spectral function $\rho _T\equiv -\frac{1}{\pi} Im\, {D^*}_R^T $.

\subsection{Interaction rate of a hard electron}

The last step of the Braaten-Pisarski method is the use of the effective Green
functions for calculating observables of hot gauge theories in the weak 
coupling limit $e\ll 1$. Famous and often discussed examples are damping or 
interaction rates of 
particles in hot relativistic plasmas (for references
see \cite{ref2}). In this section we discuss the interaction rate
of a hard electron ($p \sim T \gg eT$) in a QED plasma with zero chemical 
potential.  

The interaction rate of a massless fermion is defined by
\beq
\Gamma _{eq}(p)=-\frac{1}{2p}\, [1-n_F(p)]\> tr\, [P\sla \, Im \, 
\Sigma _R(p_0=p,{\bf p})].
\label{e29}
\eeq
The electron self energy $\Sigma $ is shown
in fig.3. The imaginary part of the diagram corresponds to the elastic scattering 
of the hard electron off thermal
electrons in the QED plasma via the exchange of a collective plasma mode. 
Since $p\gg eT$ we do not need effective 
vertices. Also, the diagram containing an effective electron propagator 
and a bare photon propagator, corresponding to Compton scattering,
can be neglected (it leads to a higher order contribution since the electron 
propagator is less singular than the photon propagator). The integral
over the photon momentum $Q$ is
dominated by small photon momenta (the Rutherford singularity). The leading 
order contribution to the interaction rate is obtained by 
integrating over the entire momentum range of the exchanged photon using 
a resummed propagator. The result is of order $e^2T$ which is greater by a 
factor of $1/e^2$ than the result one would expect from the natural two loop 
scale.  This anomalously large rate occurs because of the presence of the 
thermal photon mass in the denominator of the effective photon propagator, and 
the fact that the integral is dominated by small photon momenta.

Using the Keldysh formalism and taking the hard thermal loop limit we find, 
\beq
\Gamma _{eq}(p)=\frac{e^2T}{2\pi }\, [1-n_F(p)] \int _0^\infty dq\, q
\int _{-q}^q \frac{dq_0}{q_0}\, \left [\rho _L(Q)+\left (1-\frac{q_0^2}{q^2}
\right )\rho _T(Q)\right ],
\label{e33}
\eeq
in agreement with the result found in the ITF \cite{ref12}. 
Using the static approximation $q_0\ll q$ for the spectral functions
which is accurate to about 10\% \cite{ref2} we end up with
\beq
\Gamma _{eq}(p)\simeq \frac{e^2T}{2\pi }\, [1-n_F(p)]\, \ln \frac{const}{e},
\label{e34}
\eeq
where the $const$ under the logarithm, which comes from a singularity in
the transverse photon propagator, cannot be determined within the 
Braaten-Pisarski resummation scheme \cite{ref14}. 
Assuming an infrared cutoff of the order $e^2T$, which could be provided by the
interaction rate itself \cite{ref14}, our result (\ref{e34}) is correct to order
$e^2\ln e$. In order to determine the order $e^2$ correction one has to go beyond
the HTL resummation scheme, which lies out of the scope of the present 
investigation.

\section{Non-Equilibrium}

So far there are only a few investigations using HTL resummed Green functions 
out of equilibrium. Baier et al. \cite{ref17} have studied the photon production 
rate in chemical non-equilibrium and Le Bellac and Mabilat 
have investigated off-equilibrium reaction rates of heavy fermions in the appendix of 
Ref.\cite{ref9}.     
In this section  we want to consider a non-equilibrium situation within the
Keldysh representation by following the steps outlined in section 3.  At this point we distinguish between two separate aspects of the non-equilibrium problem.  The study of how a system that is initially out of equilibrium will relax towards equilibrium is beyond the scope of this work.  We restrict ourselves to the study of microscopic processes which take place in an out of equilibrium background, under the implicit assumption that the time scale of this microscopic process is much smaller than the time scale of the relaxation of the background towards equilibrium.  This assumption is consistent with the HTL expansion.  HTL propagators and vertices, with quasistationary distribution functions, describe the physics of modes with momenta of the order $e$ times the hard momentum scale or larger.  The damping rates which determine the relaxation time of the system are of order $e^2$ times the hard momentum scale.  Equilibration is therefore slow, at least close to equilibrium. 
 In a relativistic heavy ion collision for example, we expect a fast thermalization \cite{ref4} which could not be described by our method, and a much slower chemical equilibration \cite{ref3} where our approach should be valid \cite{ref17}. 

Out of equilibrium, difficulties arise because of the fact that (\ref{e8}) 
and (\ref{e10a}) do not hold for resummed propagators.  In equilibrium these 
relations lead  to the a priori cancellation of the pinch singularities 
associated with the product of an advanced and retarded propagator carrying
the same momentum.  Out of equilibrium, where these relations do not hold,  
the situation is more 
involved and the cancellation of pinch singularities is not automatic. We will 
discuss this problem in the remainder of
this section.

The derivation of the retarded and advanced HTL photon self energies are 
completely analogous
to the equilibrium case, because the bare electron propagator has the same
structure as in equilibrium.  
Note that the HTL approximation $|p_0|,p\ll k$ does not require the assumption
of the existence of a temperature. 
We obtain the same results for the advanced and retarded 
HTL self energies, (\ref{e17}),
(\ref{e18}) and (\ref{e20}),  with the equilibrium thermal
photon mass 
\beq
m_\gamma^2 = \frac{4e^2}{3\pi^2}\int _0^\infty dk\, k\, n_F(k)
=\frac{e^2T^2}{9} 
 \eeq
replaced by the expression 
\beq
\tilde m_\gamma ^2=\frac{4e^2}{3\pi ^2}\int _0^\infty dk\, k\, f_F(k).
\label{e35}
\eeq 
We note that there are no pinch singularities in the advanced and retarded
HTL self energies.

Since the Dyson-Schwinger equation (\ref{e21}) for the advanced and retarded
propagators is identical in equilibrium and non-equilibrium, the resummed
advanced and retarded propagators are given again by (\ref{e23})
and (\ref{e27}), using $\tilde m_\gamma$ for the thermal photon 
mass. 
We obtain the resummed symmetric propagator 
${D^*}_F^L={D^*}_{11}^L+{D^*}_{22}^L$ from the Dyson-Schwinger equation
\beq
{D^*}_{11}^L+{D^*}_{22}^L=D_{11}^L+\sum _{i,j=1}^2 D_{1i}^L\Pi _{ij}^L
{D^*}_{j1}^L+D_{22}^L+\sum _{i,j=1}^2 D_{2i}^L\Pi _{ij}^L{D^*}_{j2}^L.
\label{e36}
\eeq
Using (\ref{e5}) for the bare and full propagators and (\ref{e9}) and
(\ref{e10}) for the self energies we have, 
\beq
{D^*}_{F}^L=D_{F}^L+D_{R}^L\Pi _R^L{D^*}_{F}^L+D_F^L\Pi _{A}^L
{D^*}_{A}^L+D_{R}^L\Pi _{F}^L{D^*}_{A}^L.
\label{e37}
\eeq
 It is easy to show that this equation is solved by the
following propagator:
\bea
{D^*}_{F}^L(P)= && [1+2f_B(p_0)]\, \mbox{sgn}(p_0)\, [{D^*}_{R}^L(P)-{D^*}_{A}^L(P)]
\nonumber \\
&& +\{\Pi _F^L(P)-[1+2f_B(p_0)]\, \mbox{sgn}(p_0)\, [\Pi _R^L(P)-\Pi _A^L(P)]\}
\, {D^*}_{R}^L(P)\, {D^*}_A^L(P).
\label{e38}
\eea    
In equilibrium the second term, which might lead to pinch singularities
(because it contains the product of an advanced and a retarded propagator 
\cite{ref5}), vanishes due to (\ref{e10a}).  Equation (\ref{e24}) is 
recovered.

Out of equilibrium (\ref{e10a}) does not hold, and the second term in (\ref{e38}) does not 
automatically give zero.  We now consider this situation. A product of bare propagators in 
this expression would contain the product of delta functions which is called a pinch 
singularity:
\beq
D^L_R(P) D^L_A(P) = \frac{1}{P^2 + i {\rm sgn}(p_0)\epsilon} \,\, \frac{1}{P^2 - i {\rm sgn} 
(p_0)\epsilon} \rightarrow [\delta(P^2)]^2.
\label{pinch}
\eeq 
Consider, however, what happens when we use resummed propagators in (\ref{e38}). In this 
case we have, 
\bea
 {D^*}_R^L(P)-{D^*}_A^L(P)
&\equiv&-2\pi i \tilde \rho _L(P),\nonumber \\
{D^*}_R^L(P){D^*}_A^L(P) &=& -\pi \frac{\tilde \rho _L(P)}{Im\, \Pi _R^L(P)},
\label{xx}
\eea
where the
non-equilibrium spectral function $\tilde \rho _L$ defined in
(\ref{xx}) differs from the 
equilibrium one (\ref{e25}) only by the thermal mass (\ref{e35}).
To calculate $\Pi _F^L$ we note that (\ref{e17}) 
and (\ref{e40}) hold also out of equilibrium if we use $\tilde m_\gamma ^2$ 
in (\ref{e17}) and replace $n_F$ by $f_F$ in (\ref{e40}). Then we can write $\Pi _F^L$ as
\beq
\Pi _F^L(P)=2iA\, \frac{Im\, \Pi _R^L(P)}{p_0}, 
\label{e41}
\eeq
where the constant $A$ is given by
\beq
A=\frac{\int _0^\infty dk\, k^2\, f_F(k)\, [1-f_F(k)]}{\int _0^\infty
dk\, k\, f_F(k)}.
\label{e42}
\eeq
Inserting (\ref{xx}) and (\ref{e41}) into (\ref{e38}) and using 
\beq
\Pi _R^L(P)-\Pi _A^L(P) =2i Im\, \Pi _R^L(P)
\label{e42a}
\eeq
we obtain
\beq
{D^*}_F^L(P)=-2\pi i\, \frac{A}{p_0}\, \tilde \rho _L(P).
\label{e43} 
\eeq

In spite of  (\ref{xx}) this result holds also for a vanishing imaginary 
part of the self energy, because $Im \Pi_R^L$ drops out of the second term
of the propagator (\ref{e38}) according to (\ref{xx}), (\ref{e41}), and 
(\ref{e42a}). 

In equilibrium $A$ reduces to $2T$. Consequently (\ref{e43})
agrees with (\ref{e26}) in the equilibrium case in the soft $p_0$ (HTL) limit. For the transverse
propagator we simply have to replace $\tilde \rho _L$ by $\tilde \rho _T$ 
in (\ref{e43}).

The conclusion is the following. When (\ref{e38}) is rewritten in the form (\ref{e43}) 
it is clear that the apparent pinch singularity in the QED HTL effective photon propagator 
does not in fact occur.  Physically we have found that this singularity is regulated by the 
use of the resummed propagators.  This result agrees with that obtained by Altherr 
\cite{ref19} who found that finite results could be obtained in a scalar field theory by 
resumming pinch terms.  Since the scalar self energy has no imaginary part at one loop, 
a finite width was inserted by hand to provide the regularization.  The same result was 
also found by Baier et al. \cite{ref17} for the fermion propagator in a chemically 
non-equilibrated QCD plasma, in the HTL limit.  

We now discuss higher order calculations.  To begin we consider the `Altherr type' diagram
in the case of the electron self energy 
shown in fig.4. When this diagram is calculated with bare lines, a pinch singularity appears 
to occur because of the product of the two propagators with the same momentum dependence.  
However, resumming diagrams with all possible numbers of self energy insertions yields a 
finite result since this sum of diagrams is equivalent to calculating the one loop diagram 
with the HTL effective propagator (in the case of soft external momenta), which we have shown 
does not contain a pinch singularity.  To next order we consider the same Altherr type diagram in 
fig.4 where internal lines are HTL effective propagators, and the self energy insertion 
($\bar{\Pi}$) are the one loop diagrams with HTL effective propagators on the internal lines 
and HTL effective vertices shown in fig.5. At first glance it appears that above the light cone, where the 
imaginary part of the HTL self energy is zero, this diagram will have a pinch singularity 
that arises in the same way as for the diagram with bare propagators.  Consider, however,
what happens when we resum diagrams with all possible numbers of self energy insertions 
$\bar{\Pi}$. This procedure is equivalent to calculating the one loop diagram with an 
effective propagator $D^{**}$ that is given by the Dyson-Schwinger equation, 
\bea
{D^{**}}_{F}^L(Q)= && \frac{A}{q_0}\, [{D^{**}}_{R}^L(Q)-{D^{**}}_{A}^L(Q)]
\nonumber \\
&& +\{\bar{\Pi} _F^L(Q)-\frac{A}{q_0}\, [\bar{\Pi} _R^L(Q)-\bar{\Pi} _A^L(Q)]
\} \, {D^{**}}_{R}^L(Q)\, {D^{**}}_A^L(Q).
\label{e38b}
\eea  
The product of propagators $D_R^{**}D_A^{**}$ can be rewritten as proportional to a spectral 
function divided by the imaginary part of $\bar{\Pi}_R$ in exactly the same way as before 
(see (\ref{xx})). Thus, the regulation of the singularity will occur as before, if we can 
write the symmetric self energy $\bar{\Pi}_F$ as proportional to the imaginary part of the 
retarded self energy, as in equation (\ref{e41}). So far, this result has only been proven 
for the HTL self-energy. If it is true in general, then the mechanism outlined above for the 
HTL effective propagator will work at all orders, and all physical quantities will be free 
of pinch singularities, as expected.    

It should be noted that the self energy $\bar{\Pi}$ contains an imaginary part (damping)
also above the light cone. Hence the effective propagator $D^{**}$ has a finite width
and will regulate all pinch singularities according to Altherr \cite{ref19}. 

Quantities that are logarithmically infrared divergent using bare propagators such as
the photon production rate in a QGP can be calculated consistently to leading order
by a decomposition into a soft and a hard part \cite{ref18}. For this purpose
a separation scale $eT\ll q^*\ll T$ for the momentum $Q$ of the exchanged particle
is introduced. The hard part then follows from a two-loop self energy containing only
bare propagators analogously to fig.4. However, due to the kinematical restriction 
$-Q^2>{q^*}^2$
no pinch singularity $[\delta (Q^2)]^2$ (see (\ref{pinch})) occurs \cite{ref17}.
Hence there are no pinch singularities using the HTL resummation technique to leading order.
At higher orders a resummation beyond the HTL scheme leading to (\ref{e38b}) might
be necessary. 

Lastly, we investigate the non-equilibrium electron damping rate.  The equilibrium result  
(\ref{e33}) is modified to become, 
\beq
\Gamma _{neq}(p)=\frac{e^2}{4\pi }\, [1-f_F(p)] \int _0^\infty dq\, q
\int _{-q}^q dq_0\, \frac{A}{q_0}\, \left [\tilde \rho _L(Q)+
\left (1-\frac{q_0^2}{q^2} \right )\tilde \rho _T(Q)\right ],
\label{e44}
\eeq  
leading to the final result
\beq
\Gamma _{neq}(p)\simeq \frac{e^2A}{4\pi }\, [1-f_F(p)]\, \ln \frac{const}{e}.
\label{e45}
\eeq
The deviation of the spectral function from the equilibrium one does not 
matter here because the thermal photon mass drops out after integrating
over $q$ \cite{ref2}. Comparison with the equilibrium case (\ref{e34}) gives
\beq
\Gamma _{neq}(p)=\frac {A}{2T}\, \frac{1-f_F(p)}{1-n_F(p)}\, \Gamma _{eq}.
\label{e45a}
\eeq

Finally, we discuss the specific case of a chemically non-equilibrated 
QED plasma. Numerical transport simulations of the QGP
in relativistic heavy ion collisions show that there is rapid
thermalization in a partonic fireball. However, chemical equilibration
takes much longer, if it is achieved at all during the lifetime of the QGP
\cite{ref3,ref4}. In order to describe this deviation from chemical
equilibrium, phase space suppression factors $\lambda _{B,F}$ depending on time
-- sometimes also called 
fugacities -- are introduced \cite{ref3}. Assuming that the photons,
electrons and positrons in a QED plasma are not in chemical equilibrium, 
the distributions are given by
\bea
f_B(p_0) & = & \lambda _B\, n_B(p_0),\nonumber \\
f_F(p_0) & = & \lambda _F\, n_F(p_0),
\label{e46}
\eea
where $0<\lambda _{B,F}<1$ indicates undersaturation
and $\lambda _{B,F}>1$ oversaturation of the corresponding photons and fermions
compared to an equilibrated QED plasma. Using the distributions
(\ref{e46}) we find for the constant $A$ in (\ref{e42}) after numerical
integration $A=2T+0.192T(1-\lambda _F)$. Substitution in (\ref{e45a}) gives
\beq
\Gamma _{neq}(p)=\frac{1-\lambda _F\, n_F(p)}{1-n_F(p)}\, [1+0.096
(1-\lambda _F)]\Gamma _{eq}(p).
\label{e47}
\eeq

The non-equilibrium rate is independent of the photon fugacity $\lambda _B$ and 
depends only weakly on $\lambda _F$. This observation is the result of a cancellation
of two efects: in an
undersaturated (oversaturated) plasma the number of scattering partners is reduced
(enhanced) and, at the same time, the Debye mass $m_D^3=3m_\gamma ^3$ is reduced 
(enhanced) leading to less (more) screening. To a large extent, these two effects
cancel each other and lead to a rate that is approximately independent of
the fugacities. As a matter of fact, the non-equilibrium rate increases in an 
undersaturated plasma a little bit, because there is less Pauli blocking in this case.
In equilibrium the cancellation between the number of scattering partners (flavors) 
and the Debye screening is exact \cite{ref20}.

\section{Conclusions} 
  
In the present paper we have studied explicitly the HTL resummation 
technique in equilibrium and non-equilibrium within the RTF using 
the Keldysh representation. We have considered the HTL photon self energy, the resummed
photon propagator, and the interaction rate of a hard electron in a
QED plasma. We have pointed out the convenience of the Keldysh representation,
where only the symmetric propagators $G_F$ depend on the distribution functions and
where possible pinch terms cancel automatically in equilibrium. 

We have shown that the HTL resummation technique can be extended to 
non-equilibrium situations assuming quasistationary distributions.
This assumption does not allow us to study the equilibration of the
system; it restricts us to the study of microscopic processes taking place in an out of equilibrium background under the assumption that the time scale of this microscopic process is much smaller than the time scale of the relaxation of the background towards equilibrium. This assumption is consistent with the HTL expansion.  HTL propagators and vertices describe the physics of 
modes with momenta of the order of $e$ times the hard momentum scale or larger. The damping rates which 
determine the relaxation time of the system are of order $e^2$ times the hard momentum scale. 
Equilibration is therefore slow, at least close to equilibrium, and quasistationary 
distributions can be assumed.  In relativistic heavy ion collisions, for example, we expect a fast 
thermalization \cite{ref4}, which could not be described by our method, and a much slower chemical equilibration \cite{ref3} where our approach should be 
applicable \cite{ref17}.

The retarded and advanced HTL photon self energies in 
non-equilibrium are obtained from the equilibrium quantity by replacing the thermal mass 
of the photon by a 
non-equilibrium expression (\ref{e35}). The retarded and advanced resummed photon 
propagators have the same structure as their equilibrium counterparts.  However, 
the resummed symmetric  photon propagator 
${D^*}_F^{L,T}$ (\ref{e38}) contains an additional term (pinch term)
compared to the
equilibrium expression (\ref{e24}). This singularity is regulated by the resummed 
propagators in the pinch term in (\ref{e38}).  One obtains an expression (\ref{e43}) 
for the HTL effective propagator that has the same structure as the
equilibrium result (\ref{e26}). Therefore, there are no additional pinch singularities 
in HTL effective propagator in the non-equilibrium formalism, compared with the 
equilibrium situation. We have discussed how to extend these results beyond leading order
by an additional resummation beyond the HTL one.

Higher n-point functions could also be calculated
efficiently using the Keldysh representation \cite{ref10,ref22}. 
We expect that the absence of pinch singularities persists. Since the HTL
self energies are gauge invariant out of equilibrium (since they differ  from the
equilibrium HTL's only by the definition of the thermal masses), we expect that
Ward identities will hold out of equilibrium \cite{ref23}, and thus the structure of all HTL 
Green functions should be the same both in and out of equilibrium. 

As an example we have discussed the interaction rate of a hard electron and showed 
that the result has the same form out of equilibrium as in equilibrium.  (We note 
that this discussion of pinch singularities has no bearing on the infrared divergence 
that occurs in the HTL calculation of this quantity). We have considered a chemical 
non-equilibrium situation by
multiplying the equilibrium distribution functions by a fugacity factor.
The non-equilibrium interaction rate is approximately independent of the fugacities.

Using the formalism developed in this paper, it will be straightforward to
calculate observables in a non-equilibrium parton gas.  Examples that have already been  
considered in an equilibrated QGP include 
parton damping and transport rates, the energy loss of partons,
transport coefficients, and production rates of partons, leptons,
and photons. 
   
\acknowledgements
We would like to thank E. Braaten, P. Danielewicz, C. Greiner, U. Heinz, R. Kobes, 
S. Leupold, and B. M\"uller for stimulating and helpful discussions.

\begin{figure}
\centerline{\psfig{figure=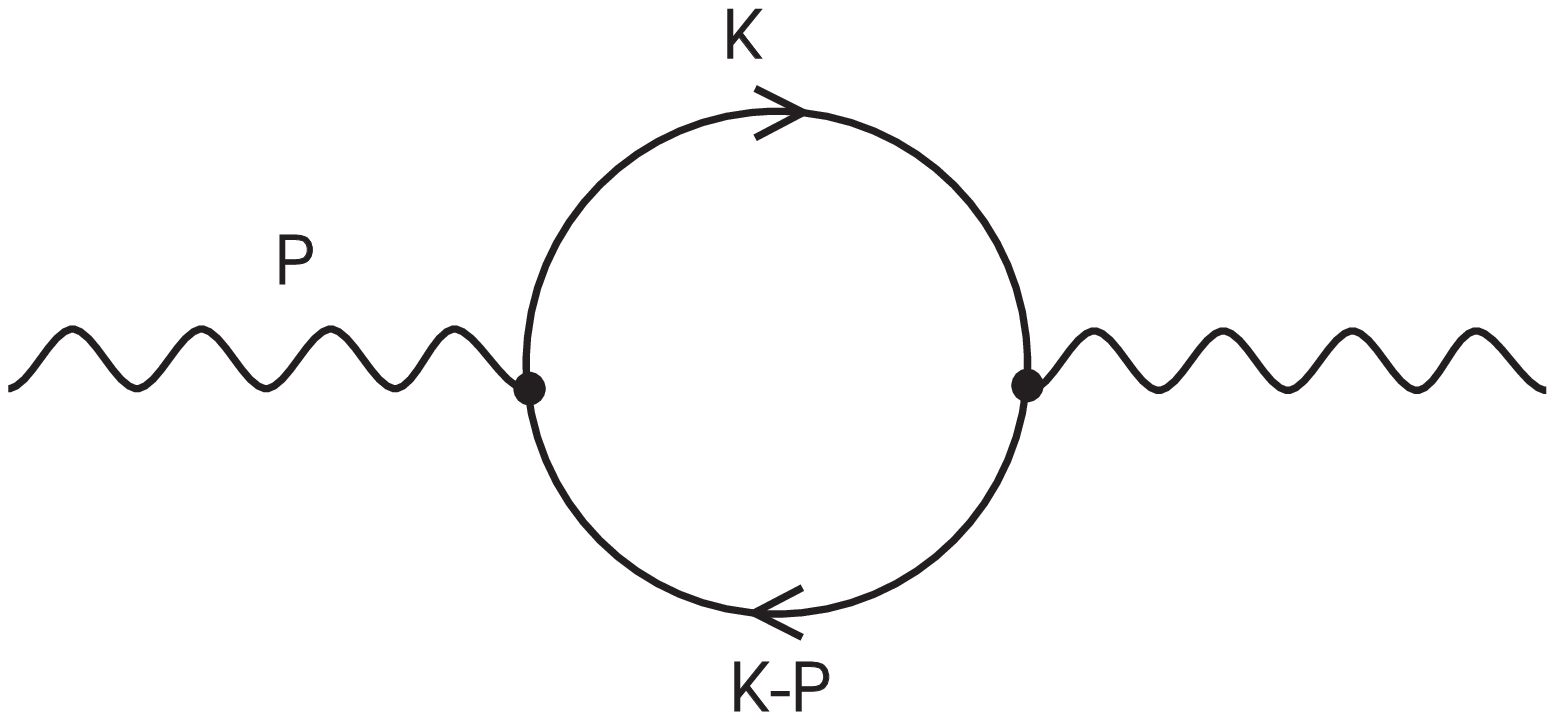,width=9cm}}
\caption{HTL photon self energy}
\end{figure}

\begin{figure}
\centerline{\psfig{figure=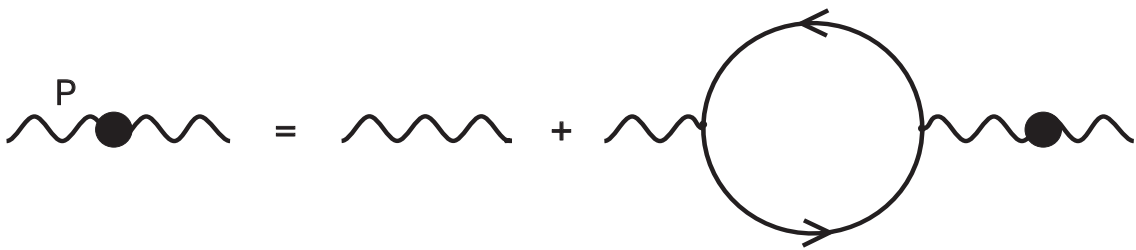,width=15cm}}
\caption{Effective photon propagator}
\end{figure}

\begin{figure}
\centerline{\psfig{figure=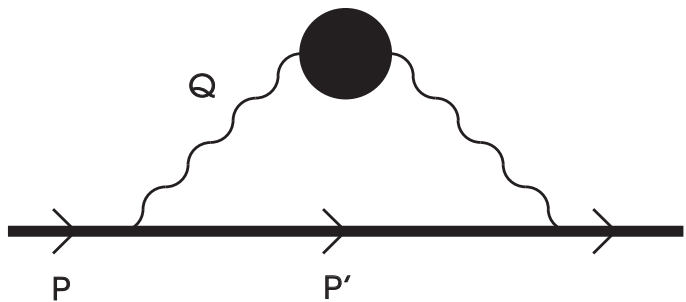,width=9cm}}
\caption{Electron self energy defining the damping rate of a hard electron}
\end{figure}

\begin{figure}
\centerline{\psfig{figure=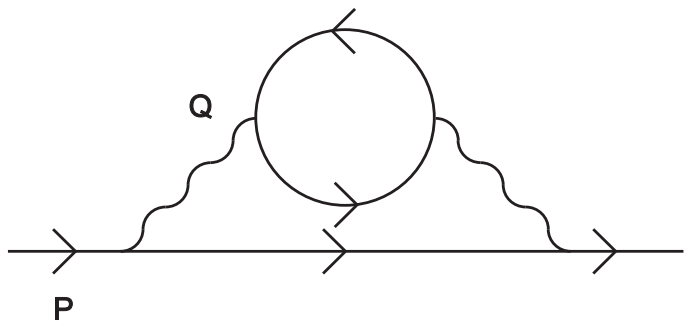,width=9cm}}
\caption{Electron self energy containing a pinch singularity}
\end{figure}

\begin{figure}
\centerline{\psfig{figure=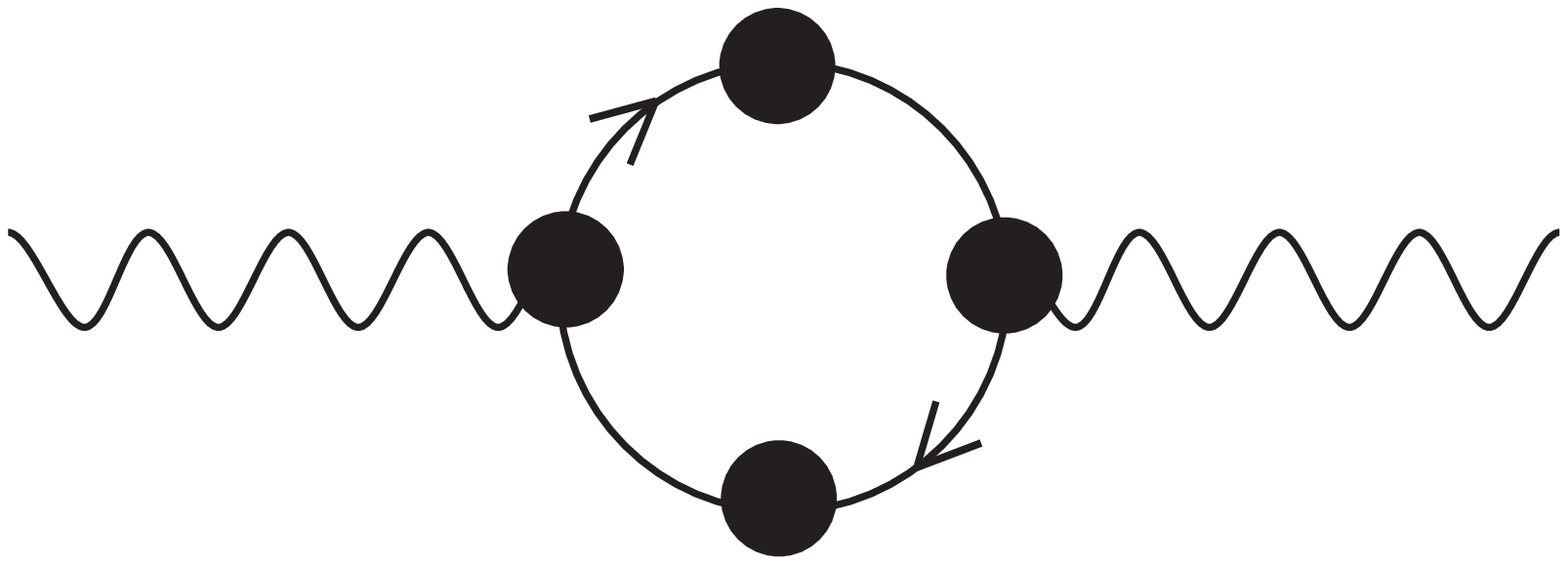,width=9cm}}
\caption{Photon self energy $\bar \Pi $}
\end{figure}


\begin{references}

\bibitem{ref1} E. Braaten and R.D. Pisarski, Nucl. Phys. {\bf B337}, 569 (1990).

\bibitem{ref2} M.H. Thoma, in {\it Quark-Gluon Plasma 2}, edited by R. Hwa 
(World Scientific, Singapore 1995), p.51.

\bibitem{ref3} T.S. Bir\'o, E. van Doorn, B. M\"uller, M.H. Thoma, and X.N. 
Wang, Phys. Rev. C {\bf 48}, 1275 (1993). 

\bibitem{ref4} K. Geiger, Phys. Rep. {\bf 258}, 238 (1995); X.N. Wang, Phys. Rep.
{\bf 280}, 287 (1997).

\bibitem{ref5} N.P. Landsmann and C.G. van Weert, Phys. Rep. {\bf 145}, 141 (1987).

\bibitem{ref6} K. Chou, Z. Su, B. Hao, and L. Yu, Phys. Rep. {\bf 118}, 1 (1985).

\bibitem{ref7} L.V. Keldysh, JETP {\bf 20}, 1018 (1965). 

\bibitem{ref8} T. Altherr and D. Seibert, Phys. Lett. B {\bf 333}, 149 (1994).


\bibitem{ref10} M.E. Carrington and U. Heinz, Eur. Phys. J. C {\bf 1}, 619 
(1998).


\bibitem{ref11b} A. Peshier, K. Schertler, and M.H. Thoma, hep-ph/9708434, 
Ann. Phys. (N.Y.) (in press).

\bibitem{ref11c} H.A. Weldon, Phys. Rev. D {\bf 26}, 1394 (1982); V.V. Klimov, Sov. Phys. 
JETP {\bf 55}, 199 (1982).

\bibitem{ref11a} R.D. Pisarski, Physica A {\bf 158}, 146 (1989).

\bibitem{ref12} E. Braaten and M.H. Thoma, Phys. Rev. D {\bf 44}, 1298 (1991).


\bibitem{ref14} J.P. Blaizot and E. Iancu, Phys. Rev. Lett. {\bf 76}, 3080 (1996).



  
\bibitem{ref17} R. Baier, M. Dirks, K. Redlich, and D. Schiff, Phys. Rev. D {\bf 56} 
(1997) 2548.

\bibitem{ref9} M. Le Bellac and H. Mabilat, Z. Phys. C {\bf 75}, 137 (1997).

\bibitem{ref18a} P.A. Henning, Nucl. Phys. {\bf A582}, 633 (1994); P.F. Bedaque, 
Phys. Lett. B {\bf 344}, 23 (1995); C. Greiner and S. Leupold, hep-ph/9802312
and hep-ph/9804239.

\bibitem{ref19} T. Altherr, Phys. Lett. B {\bf 341}, 325 (1995).

\bibitem{ref18} E. Braaten and T.C. Yuan, Phys. Rev. Lett. {\bf 66}, 2183 (1991).

\bibitem{ref20} M.H. Thoma, Phys. Rev. D {\bf 49}, 451 (1994).

\bibitem{ref22} H. Defu and U. Heinz, hep-ph/9704392, Eur. Phys. J. C 
(in press).

\bibitem{ref23} M.E. Carrington, H. Defu, and M.H. Thoma, hep-ph/9801103.

\end{references}
\end{document}